\documentclass[preprint,titlepage,showpacs,preprintnumbers,amsmath,amssymb,noshowkeys,superscriptaddress]{revtex4}
\usepackage{graphicx}
\usepackage{rotating}
\usepackage{epsfig}
\usepackage{color}
\usepackage{array}
\usepackage[dvipsnames]{xcolor}
\usepackage[utf8]{inputenc}
\usepackage[T2A]{fontenc}

\begin{document}

\renewcommand{\abstractname}{Abstract}
\renewcommand{\figurename}{Fig.}
\renewcommand{\tablename}{Table.}
\newcommand{\Ast}{\mathop{\scalebox{0.8}{\raisebox{0.35ex}{$\ast$}}}}
\renewcommand{\andname}{\ignorespaces}

\title{Cumulative protons and pions at the intermediate energies in the FRAGM experiment}

\author{M.A.~Martemianov}
\email[E-mail: ]{mmartemi@gmail.com}
\affiliation{NRC «Kurchatov Institute», Moscow, Russia}%

\author{B.M.~Abramov}
\affiliation{NRC «Kurchatov Institute», Moscow, Russia}%

\author{S.A.~Bulychjov}
\affiliation{NRC «Kurchatov Institute», Moscow, Russia}%

\author{I.A.~Dukhovskoy}
\affiliation{NRC «Kurchatov Institute», Moscow, Russia}%

\author{V.V.~Kulikov}
\affiliation{NRC «Kurchatov Institute», Moscow, Russia}%

\author{A.A.~Kulikovskaya}
\affiliation{NRC «Kurchatov Institute», Moscow, Russia}
\affiliation{EANPO HE <<Moscow Technological Institute>>, Moscow, Russia}

\author{M.A.~Matsyuk}
\affiliation{NRC «Kurchatov Institute», Moscow, Russia}%

\begin{abstract}
One of the important research objectives of the FRAGM experiment operating at the TWAC heavy-ion accelerator-storage complex was the study of particle production in the cumulative region. Differential cross sections were measured for the production of pions -- resulting from the fragmentation of carbon nuclei at an energy of 3.2~GeV/nucleon, as well as protons with energy range from 300 to 2000~MeV/nucleon at an angle of 3.5$^\circ$ relative to the incident ion beam. The parameterization of cumulative effect used in experiments investigating nucleus-nucleus interactions is discussed. It is shown, that the energy dependencies of these parameters are analogous to those observed in nucleon-nucleus interactions and indicate the presence of nuclear scaling behavior.
\end{abstract}

\pacs{25.70.Mn, 25.70.Pq}
\maketitle

\section{Introduction}

At present there exist several theoretical descriptions of the mechanism of cumulative processes in nucleus-nucleus interactions. The particle production, kinematically forbidden in free nucleon interactions, is explained by the presence of multiquark clusters, density fluctuations of nuclear matter, and also by the mechanism of multiple rescattering~\cite{PaperIntr01,PaperIntr02,PaperIntr03}. The most complete experimental data on cumulative processes were obtained in proton-nucleus experiments. The actively developing study of nucleus-nucleus interactions at present allows one to hope for a deepening of our knowledge of hadron interactions in nuclear matter and, in particular, of cumulative processes.

For the most general phenomenological description of the energy spectrum of cumulative particles, a sum of two exponentials is often used, and the invariant cross section can be represented as:
\begin{center}\vspace{-0.2cm}$Ed^3\sigma/d^3p$~$\sim$~$A_S\hspace{0.2em}\text{exp}(-T/T_S)$ + $A_C\hspace{0.2em}\text{exp}(-T/T_C)$,\vspace{-0.2cm}\end{center}
where $T$ is the kinetic energy of the fragment, $A_S$ and $A_C$ are normalization constants for the statistical term describing the spectrum of evaporation nucleons and for the cumulative term, and $T_S$ and $T_C$ are the corresponding slope parameters~\cite{TempParameters1,TempParameters2}. This approximation can serve as a fair approximation to the energy spectra of both nucleons and heavier fragments.

\section{Proton momentum spectrum}

The FRAGM experiment~\cite{PaperFRAGM} was performed at the TWAC heavy-ion accelerator storage complex~\cite{PaperTWAC}, designed for accelerate ions from 0.3 to 4.0~GeV/nucleon. The experimental setup included a two-stage magneto-optical channel that captured fragments of interactions of accelerator beam ions with an internal target, producing at an angle of 3.5$^\circ$. Fragments were registered by a system of scintillation detectors and identified by measuring ionization losses (fragment charge function) and time of flight (fragment mass number function). Fragment momentum spectra were measured by scanning the magneto-optical channel rigidity with a step of 50--100~MeV/{\it c} and normalized to the monitor readings of a monitor, which for calculating fragment production cross sections was normalized to known proton production cross section at the fragmentation peak maximum.
\begin{figure}[hbt!]
\includegraphics[scale=0.65]{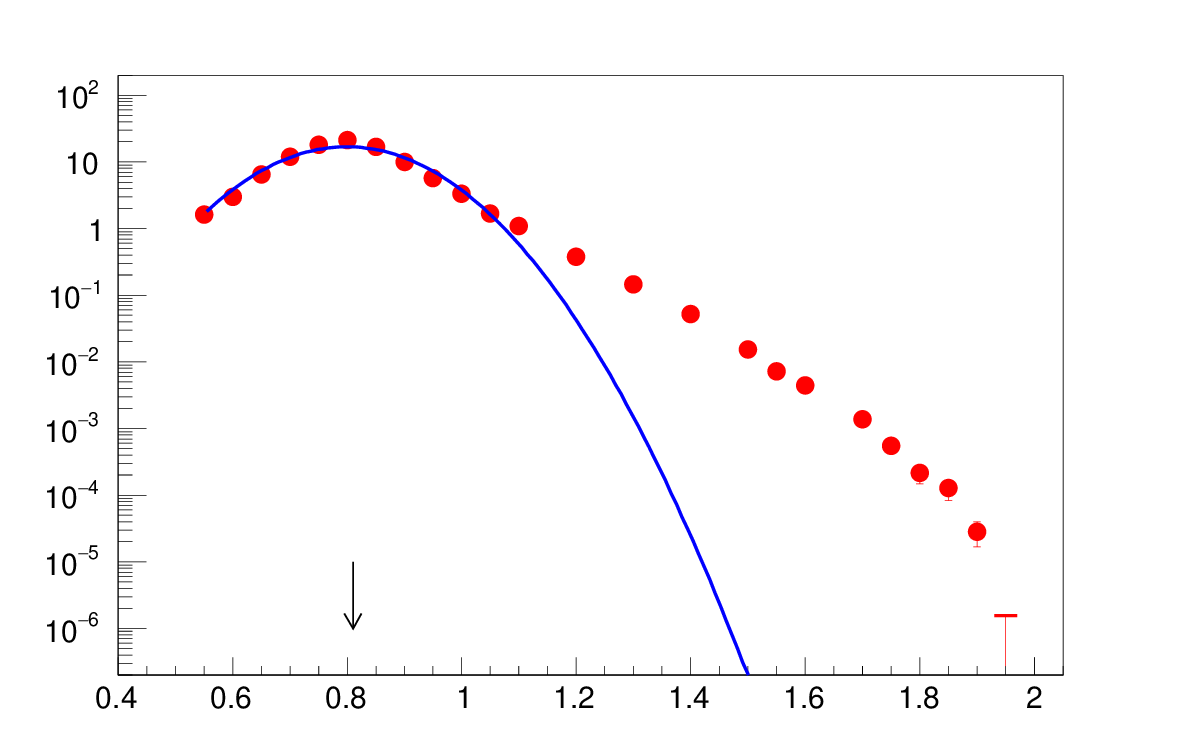}
\setlength{\unitlength}{5.0pt} 
\begin{picture}(100,0)
\put(18,47.8){\fontsize{12}{12}\selectfont{{\it d}$^{\hspace{0.1em}2}\sigma/$({\it dpd}$\Omega$),
mb$/$(MeV$/${\it{c}}\hspace{0.1cm}$\cdot$\hspace{-0.1cm} sr)}}
\put(67.,4.5){\fontsize{12}{12}\selectfont{$p$, GeV/{\it c}}}
\put(34.0,17.0){\fontsize{14}{14}\selectfont{$p_{0}$}}
\end{picture}\vspace{-1.2cm}	
\caption{Differential cross section for proton yield in the fragmentation of carbon nuclei at 
$T_0$~=~300~MeV/nucleon measured in the FRAGM experiment. The arrow marks the position of fragmentation peak maximum $p_0$. At a proton momentum of 1.95 GeV/{\it c} the upper cross section limit is indicated. The line shows fitting of the cross section by a Gaussian function in the maximum region.}\label{Pict01}
\end{figure}

The differential cross section $d^2\sigma/(dpd\Omega)$ for proton yield in the fragmentation of carbon nuclei at 300~MeV/nucleon on a beryllium target versus momentum is shown in Fig.~\ref{Pict01}. The data cover six orders of magnitude of cross section. The peak of this distribution is located in the region 0.8~GeV/{\it c}, corresponding to the momentum per nucleon of the incident nucleus. Near the maximum the peak is well described by a Gaussian distribution. Extrapolation of this description to a wider region of proton momenta indicates a large deviation from Gaussian dependence and a transition to exponential dependence $\text{exp}(-x/x_0)$, typical of cumulative processes, where $x=p/p_0$ is the ratio of the proton momentum to the momentum per nucleon of the incident nucleus, and $x_0$ is the cumulative parameter equal to 0.105. With increasing momentum this dependence will be limited by energy conservation.

At high energies of the incident nucleus, due to the small cross section, reaching this limit is problematic. However, in our experiment at a kinetic energy of the incident carbon nucleus of 300~MeV/nucleon it was possible to cross the kinematic boundary of proton yield in the reaction $^{12}$C~+~$p$~$\rightarrow$~$p$~+~$X$, which is 1.86~GeV/{\it c} or $x=2.3$. This is actually quasi-elastic proton scattering backward on a carbon nucleus. 
\begin{figure}
\includegraphics[scale=0.65]{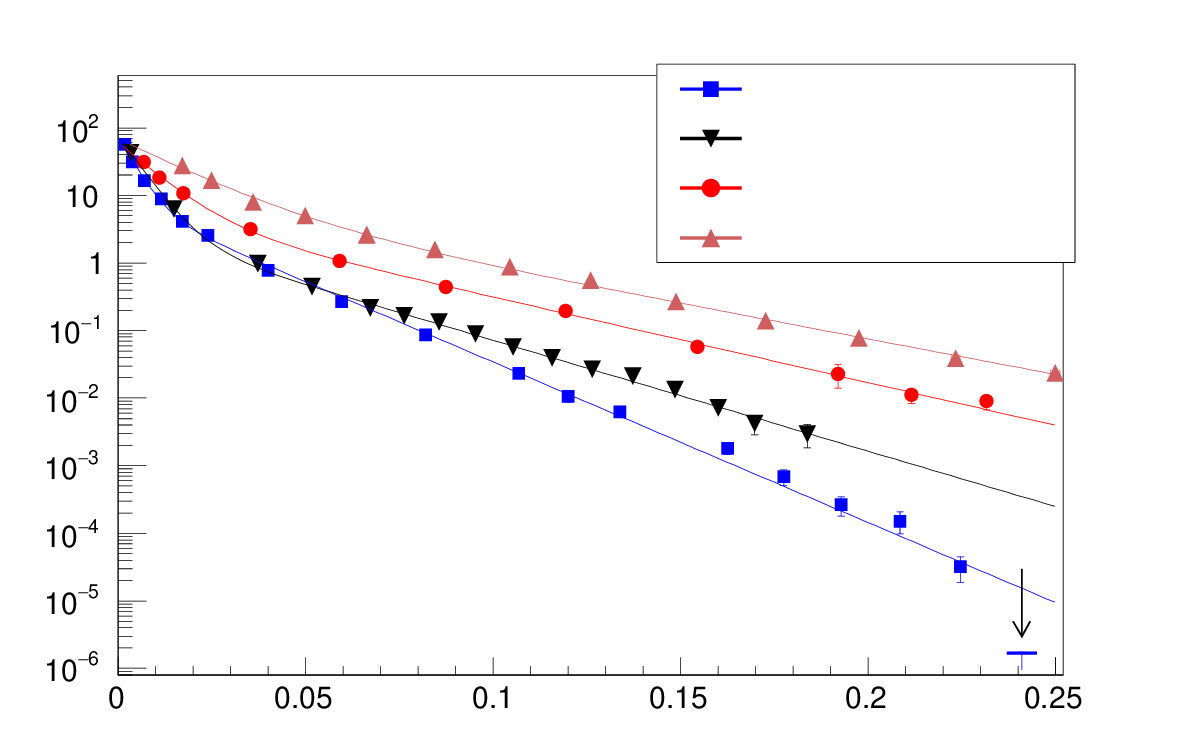}
\setlength{\unitlength}{5.0pt}
\begin{picture}(100,0)
\put(18,47.8){\fontsize{11}{11}\selectfont{{\it Ed}$^{\hspace{0.1em}2}\sigma/$({\it p$^2$dpd}$\Omega$), b$/$(GeV$^2$$/${\it{c$^3$}}\hspace{0.1cm}$\cdot$\hspace{-0.1cm} sr)}}
\put(66.5,4.5){\fontsize{12}{12}\selectfont{$T_{RF}$, GeV}}
\put(57.6,45.2){\fontsize{10}{10}\selectfont{$T_0$\hspace{0.2em}=\hspace{0.2em}300 MeV/nucl.}}
\put(57.6,42.1){\fontsize{10}{10}\selectfont{$T_0$\hspace{0.2em}=\hspace{0.2em}600 MeV/nucl.}}
\put(57.6,39.0){\fontsize{10}{10}\selectfont{$T_0$\hspace{0.2em}=\hspace{0.2em}950 MeV/nucl.}}
\put(57.6,35.8){\fontsize{10}{10}\selectfont{$T_0$\hspace{0.2em}=\hspace{0.2em}2000 MeV/nucl.}}
\put(54.5,33.0){\color{blue}\fontsize{14}{14}\selectfont{}}
\put(63.2,30.0){\color{blue}\fontsize{14}{14}\selectfont{}}
\end{picture}\vspace{-1.2cm}
\caption{Invariant cross sections for proton yield versus kinetic energy in the rest system of incident nucleus in carbon fragmentation at energies from 300 to 2000 MeV/nucleon. The arrow indicates the upper cross section boundary at $T_0$ = 300 MeV/nucleon.}\label{Pict02}
\end{figure}

Above this boundary a sharp drop of the cross section is observed, and at a momentum of 1.95~GeV/{\it c} not a single event of this reaction was registered. Fig.~\ref{Pict01} shows an upper limit indicating that at the 90 \% confidence level the cross section is less than this value. Exceeding the kinematic boundary of elastic proton scattering on a carbon nucleus is associated with the fact that proton is not free but bound in the beryllium nucleus. Internal motion of nucleons in the nucleus leads to smearing of the above mentioned boundary. However, the behavior of the invariant cross section both near and beyond this boundary has not yet been discussed, as well as over the entire region up to the kinematic limit of proton momentum in the reaction $^{12}$C~+~$^{9}$Be~$\rightarrow$~$p$~+~$X$, which is 3.38~GeV/{\it c}, almost twice our measurements.

\begin{figure}\vspace{1.0cm}
\includegraphics[scale=0.65]{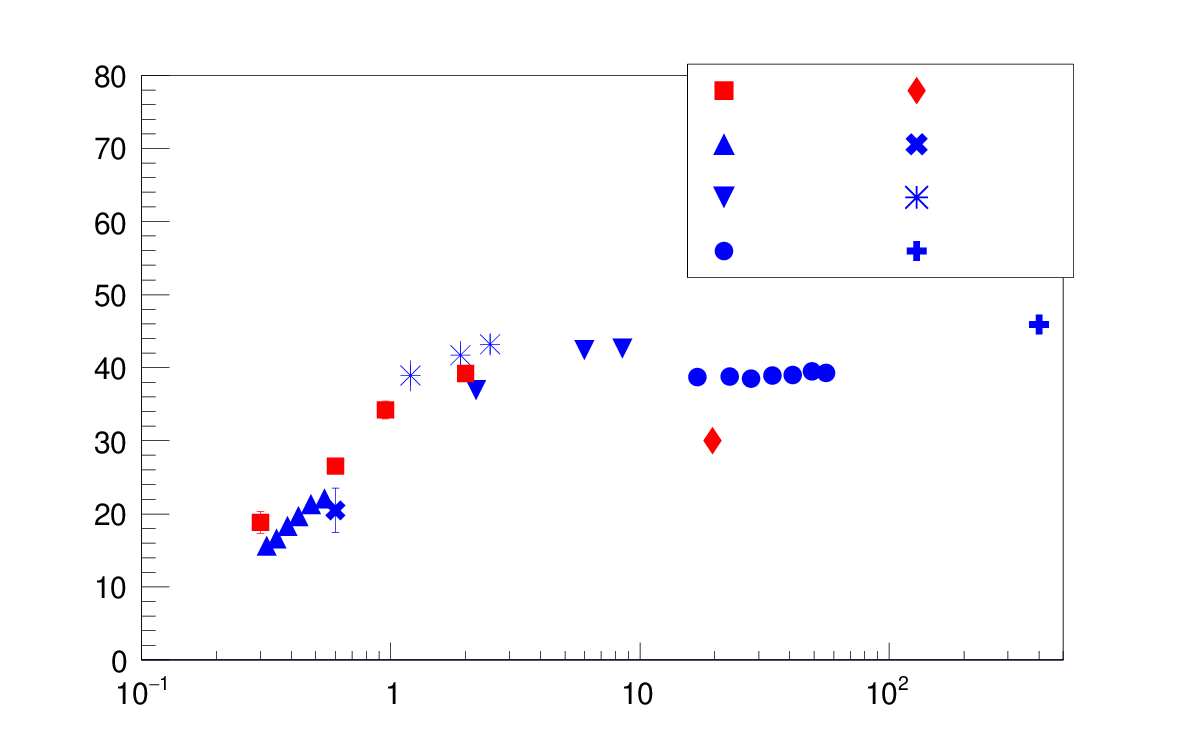}
\setlength{\unitlength}{5.0pt}
\begin{picture}(100,0)
\put(20.,47.8){\fontsize{12}{12}\selectfont{$T_C$, MeV}}
\put(60.,4.5){\fontsize{12}{12}\selectfont{$T_0$, GeV/nucleon}}
\put(57.2,44.8){\fontsize{10}{10}\selectfont{FRAGM}}
\put(69.4,44.8){\fontsize{10}{10}\selectfont{FODS}}
\put(57.2,41.6){\fontsize{10}{10}\selectfont{SIN}}
\put(69.4,41.6){\fontsize{10}{10}\selectfont{SREL}}
\put(57.2,38.3){\fontsize{10}{10}\selectfont{ITEP}}
\put(69.4,38.3){\fontsize{10}{10}\selectfont{COSY}}
\put(57.2,34.9){\fontsize{10}{10}\selectfont{IHEP}}
\put(69.4,34.9){\fontsize{10}{10}\selectfont{FNAL}}
\end{picture}\vspace{-1.0cm}
\caption{$T_C$ parameters for protons obtained in the FRAGM experiment~\cite{PaperFRAGM300} and in the FODS experiment at IHEP~\cite{Data_AA} in carbon fragmentation at 19.6 GeV/nucleon, as well as in nucleon-nucleus interactions SIN~\cite{Data_Pr01}, SREL~\cite{Data_Pr02}, ITEP~\cite{Data_Pr03}, COSY~\cite{Data_Pr04}, IHEP~\cite{Data_Pr05} and FNAL~\cite{Data_Pr06}.}\label{Pict03}
\end{figure}
The distribution of the invariant cross section $Ed^2\sigma/(p^2dpd\Omega)$ for proton yield versus their kinetic energy is shown in Fig.~\ref{Pict02}. The data are given for fragmentation of carbon nuclei with initial energy from 300 to 2000~MeV/nucleon~\cite{PaperFRAGM300, PaperFRAGM600,PaperFRAGM950,PaperFRAGM2000} and are well described by the sum of two exponential functions, where the parameter $T_C$ characterizes the cross section decrease in the cumulative region. The $T_C$ dependence for proton yield in nucleus-nucleus interactions at different incident nucleus energies is shown in Fig.~\ref{Pict03} together with similar data for proton-nucleus interactions. It is seen that the FRAGM experiment data demonstrate growth of the $T_C$ parameter in the energy range from 300 to 2000~MeV/nucleon, similar to that observed in proton-nucleus interactions. The single measurement at 19.6~GeV/nucleon performed at IHEP~\cite{Data_AA} is an indication of the onset of scaling behavior of this parameter in nucleus-nucleus interactions, reliably established in proton-nucleus interactions~\cite{Data_Pr01,Data_Pr02,Data_Pr03,Data_Pr04}.

\section{Cumulative pions}

Experimental data on charged pion yields were measured in the fragmentation of carbon nuclei on a beryllium target at 3.2~GeV/nucleon~\cite{PaperPions}. For $\pi^+$ identification a Cherenkov counter was used, which allowed suppression of the proton background after selecting a signal in the analysis 
\begin{figure}
\includegraphics[scale=0.65]{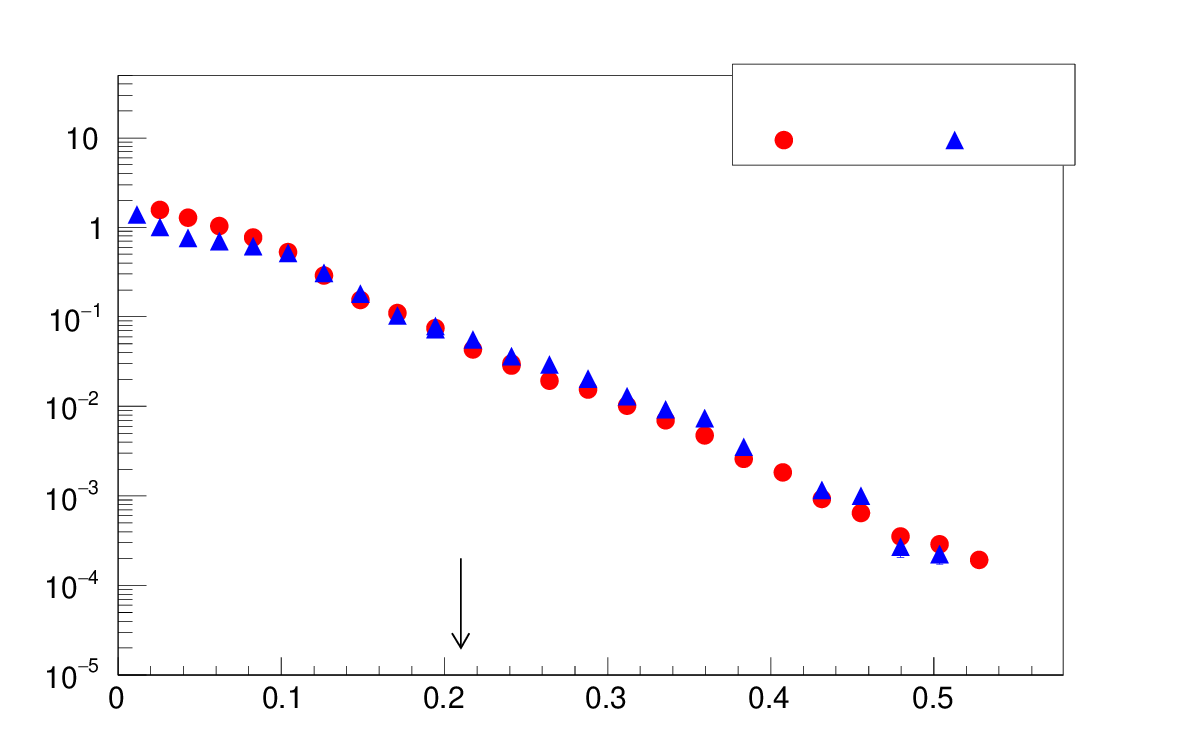}
\setlength{\unitlength}{5.0pt}
\begin{picture}(100,0)
\put(18,47.8){\fontsize{12}{12}\selectfont{{\it Ed}$^{\hspace{0.1em}2}\sigma/$({\it p$^2$dpd}$\Omega$),
b$/$(GeV$^2$$/${\it{c$^3$}}\hspace{0.1cm}$\cdot$\hspace{-0.1cm} sr)}}
\put(67.,4.5){\fontsize{12}{12}\selectfont{$T_{RF}$, GeV}}
\put(57.8,44.7){\fontsize{11}{11}\selectfont{$T_0$\hspace{0.1em}=\hspace{0.1em}3.2 GeV/nucl.}}
\put(62.6,42){\fontsize{12}{12}\selectfont{$\pi^+$}}
\put(73.0,42){\fontsize{12}{12}\selectfont{$\pi^-$}}
\end{picture}\vspace{-1.3cm}
\caption{Invariant cross sections for charged pion yield as a function of kinetic energy in the rest system of incident nucleus in carbon fragmentation at 3.2 GeV/nucleon~\cite{PaperPions}. The arrow marks the boundary of cumulative region.}\label{Pict04}
\end{figure}
of amplitude--time-of-flight correlation distributions. The invariant cross section distribution for pion yield versus the pion kinetic energy in the rest system of the incident carbon nucleus $T_{RF}$ is shown in Fig.~\ref{Pict04}. The $\pi^+$ and $\pi^-$ yields are practically identical. Pion spectra are well described by an exponential dependence, where the slope parameter $T_C$ is close to 50~MeV. It is also seen that the transition to the cumulative region hardly affects the magnitude of $T_C$, in contrast to proton production.

\begin{figure}
\vspace{-1.5cm}
\includegraphics[scale=0.65]{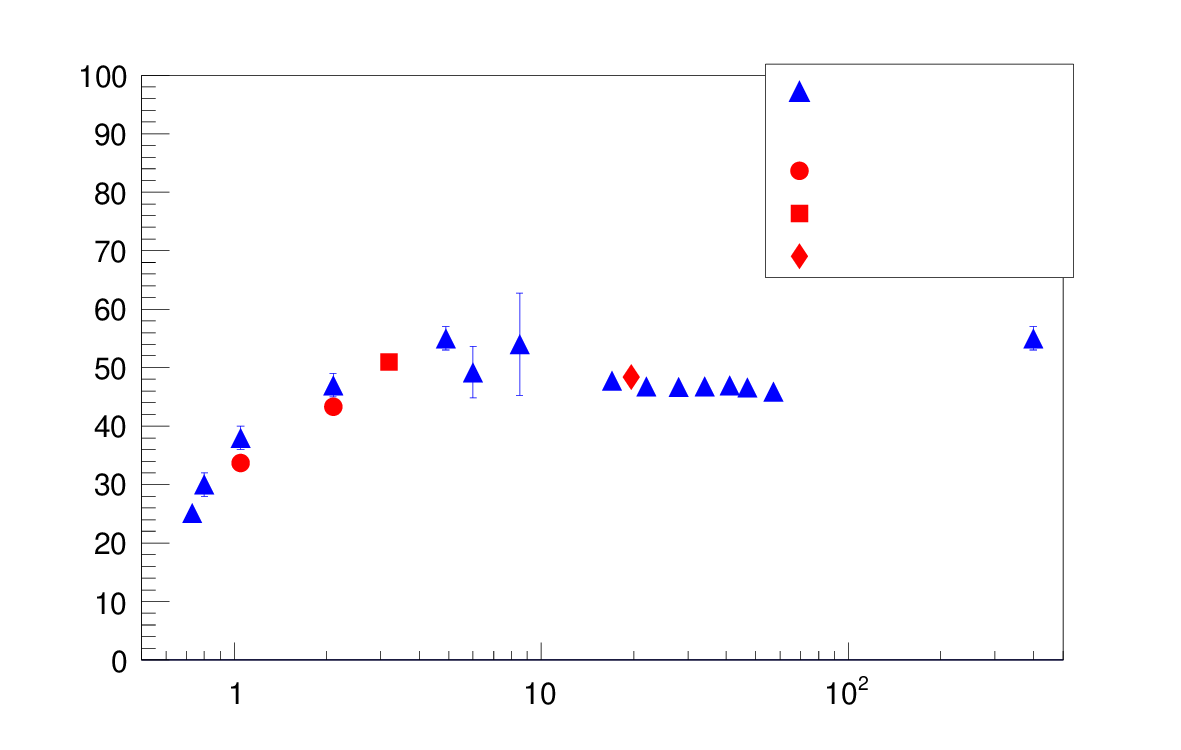}
\setlength{\unitlength}{5.0pt}
\begin{picture}(100,0)
\put(20.,47.8){\fontsize{12}{12}\selectfont{$T_C$, MeV}}
\put(60.,4.5){\fontsize{12}{12}\selectfont{$T_0$, GeV/nucleon}}
\put(61.8,45.2){\fontsize{12}{12}\selectfont{pA\hspace{0.2em}--\hspace{0.3em}reactions}}
\put(61.3,42.4){\fontsize{12}{12}\selectfont{AA\hspace{0.2em}--\hspace{0.3em}reactions}}
\put(61.8,39.9){\fontsize{10}{10}\selectfont{LBL}}
\put(61.8,37.2){\fontsize{10}{10}\selectfont{FRAGM}}
\put(61.8,34.6){\fontsize{10}{10}\selectfont{FODS}}
\end{picture}\vspace{-1.0cm}
\caption{Dependence of the $T_C$ parameter for $\pi$-mesons on the kinetic energy of the incident nucleus for proton-nucleus reactions (data for $T_0$ at 0.73~GeV~\cite{Data_pA01}, 0.8--4.89~GeV~\cite{Data_pA02,Data_pA03}, 6.0 and 8.5~GeV~\cite{Data_pA04}, 15--65~GeV~\cite{Data_pA05} and 400~GeV~\cite{Data_pA06}) and for carbon fragmentation on carbon and beryllium targets ($T_0$ at 1.05 and 2.1~GeV/nucleon obtained at LBL~\cite{Data_CA01}; 3.2~GeV/nucleon and 19.6~GeV/nucleon from the FRAGM~\cite{PaperPions} and FODS~\cite{Data_AA} experiments, respectively).}\label{Pict05}
\end{figure}

The results of $T_C$ measurements obtained in the FRAGM experiment supplement existing data on the study of cumulative pions and protons in nucleus-nucleus interactions. Experimental data on $T_C$ versus the kinetic energy of the incident nucleus for proton-nucleus and nucleus-nucleus interactions are shown in Fig.~\ref{Pict05}. The behavior of this parameter for proton-nucleus reactions demonstrates an increase in the energy range from 700~MeV to several~GeV, as well as a plateau at high energies, which is called nuclear scaling~\cite{Data_pA01,Data_pA02,Data_pA03,Data_pA04,Data_pA05, Data_pA06}. Nucleus-nucleus interactions exhibit similar behavior of the parameter $T_C$. It increases from 33 to 51~MeV in the energy range from 1 to 3.2~GeV/nucleon~\cite{Data_CA01}. The measurement at maximum energy was performed in the FODS experiment in carbon-carbon interactions at 19.6~GeV/nucleon, with value $T_C~=~48.4~\pm~0.7$~MeV. This result indicates scaling behavior of the cumulative temperature parameter in nucleus-nucleus interactions.

\section{Conclusion}

In the FRAGM experiment the energy spectra of pion and proton yield in the cumulative region were measured in the fragmentation of carbon nuclei on a beryllium target in the intermediate energy range. An estimate of the cumulative effect magnitude was made within the phenomenological description of the energy spectrum of cumulative particles as a sum of two exponential functions. A comparison of the $T_C$ parameter with data from other experiments is given. The FRAGM data significantly supplement world data and demonstrate an increase of this parameter in the intermediate energy range of the incident nucleus, close to that observed in nucleon-nucleus interactions.

\section*{Acknowledgments}\vspace{-0.5cm}
The authors are grateful to the staff of the TWAC accelerator complex and the technical staff of the FRAGM experiment for their great contribution to the measurements.

\section*{Funding}\vspace{-0.5cm}
The work was performed within the state assignment of NRC~<<Kurchatov Institute>>.

\end{document}